\documentclass[a4paper]{jpconf}

\usepackage{graphicx}
\usepackage{amsmath}
\usepackage{bm}
\usepackage{mathrsfs}

\usepackage{url}

\usepackage{color}

\begin{document}

%%%%%%%%%%%%%%%%%%%%%%%%Front Matter%%%%%%%%%%%%%%%%%%%%%%%%%%%%%%%%%%
%%%%%%%%%%%%%%%%%%%%%%%%%%%%%%%%%%%%%%%%%%%%%%%%%%%%%%%%%%%%%%%%%%%%%%

\title{
CGC initial conditions at RHIC and LHC
}
%\date{\today}

\author{Javier L. Albacete$^1$, Adrian Dumitru$^{2,3}$, Yasushi Nara$^4$}

\address{$^1$ Institut de Physique Th\'eorique - CEA/Saclay, 91191 Gif-sur-Yvette cedex, France.}

\address{$^2$ Department of Natural Sciences, Baruch College, CUNY,
17 Lexington Avenue, New York, NY 10010, USA}

\address{$^3$ RIKEN BNL Research Center, Brookhaven National
Laboratory, Upton, NY 11973, USA }

\address{$^4$ Akita International University, Yuwa, Akita-city 010-1292, Japan}

\ead{javier.lopez-albacete@cea.fr}
\ead{Adrian.Dumitru@baruch.cuny.edu}
\ead{nara@aiu.ac.jp}

%\maketitle

\begin{abstract}
Monte-Carlo implementations of $k_T$-factorization formula
with both KLN and running-coupling BK unintegrated gluon distributions
for nucleus-nucleus collisions are used to analyze recent experimental
data on the particle multiplicities
from RHIC(Au+Au@200GeV) and LHC(Pb+Pb@2.76TeV).
We also compare the predicted transverse energy at midrapidity to new
data from ALICE.
\end{abstract}

\section{Introduction}

Relativistic Heavy Ion Collider (RHIC) at BNL
has started in 2000 and many interesting data have been obtained.
Experiments at Large Hadron Collider (LHC) at CERN has just started.
It is well known fact that nearly perfect fluid picture works
to explain large elliptic flow discovered at RHIC and LHC.
However, there is no detailed understanding on the particle production
in high energy hadronic collisions. Especially subsequent
non-equilibrium evolution of the system (Glasma) created in nuclear collisions
is not known well. Without knowing the detailed glue dynamics in early 
stages of nucleus collision, correct initial conditions for hydrodynamics
would not be obtained.
First important issue is to understand the gluon production
in the first moment of collision.
In order to study such gluon production in nucleus-nucleus collision,
we shall use the Monte-Carlo implementation of $k_t$-factorization formulation
in which fluctuations of the position of nucleon inside a nucleus
is taken into account, and we can study nucleus-nucleus collision
event-by-event.

\section{Theoretical Models}

We will use the Monte-Carlo KLN (MC-KLN) model~\cite{MCKLN} and its extension
to running coupling Balitsky-Kovchegov (MCrcBK)~\cite{mckt}
for the computation of gluon production in heavy ion collisions.
Both model apply the $k_t$-factorized formula~\cite{KLN}
in the transverse plane perpendicular to the beam axis locally.
The number distribution of produced gluons is given by
\begin{equation}
  \frac{dN_g}{d^2 r_{\perp}dy} \sim \kappa_g
   \frac{4N_c}{N_c^2-1}
    \int%^{p_\perp^\mathrm{max}}
    \frac{d^2p_\perp}{p^2_\perp}
      \int d^2k_\perp \;\alpha_s\,
          \phi_A(x_1,\bm{k}_\perp^2)\;
       \phi_B(x_2,(\bm{p}_\perp{-}\bm{k}_\perp)^2)~,
      \label{eq:ktfac}
\end{equation}
with $N_c=3$ the number of colors.
Here, $p_\perp$ and $y$ denote the
transverse momentum and the rapidity of the produced gluons, respectively. 
The light-cone momentum fractions of the colliding gluon
ladders are then given by $x_{1,2} = p_\perp\exp(\pm y)/\sqrt{s_{NN}}$,
where $\sqrt{s_{NN}}$ denotes the center of mass energy.
A constant gluon multiplication factor $\kappa_g=5$ is assumed in the
MCrcBK model to obtain $p_\perp$-integrated hadron yields~\cite{mckt}. 
It does not appear in the corresponding formula for the transverse
energy $dE_\perp/dy$ since gluon splitting and hadronization
conserves energy.

At each grid point, we compute the thickness function $T_{A}(\bm{r}_\perp)$
to obtain the local saturation scale and then compute gluon production
probability. Within a hard disk nucleon approximation,
thinkness function is given by
\begin{equation}
T_{A}(\bm{r}_\perp) = \frac{\text{number of nucleon within $S$}}{S}
\end{equation}
where we assume that the area $S$ has the same value as the inelastic
proton-proton cross section at the incident energy of $\sqrt{s_{NN}}=200$
GeV independent colliding energy.

In MC-KLN model, saturation momentum is defined as
\begin{equation}
Q_{s,A}^2 (x; \bm{r}_\perp)  =  2\ \text{GeV}^2
\frac{T_A(\bm{x}_\perp)}{1.53\ \text{fm}^{-2}}
\left(\frac{0.01}{x}\right)^{\lambda} \ ,
\label{eq:qs2}
\end{equation}
where $\lambda$ is a free parameter which is expected
to have the range of $0.2<\lambda<0.3$ from HERA global analysis.
In MC-KLN, we assume the gluon distribution
function as
\begin{equation}
\label{eq:uninteg}
  \phi_{A,B}(x,k_\perp^2;\bm{r}_\perp)\sim
    \frac{1}{\alpha_s(Q^2_s)}\frac{Q_s^2}
       {{\rm max}(Q_s^2,k_\perp^2)}~,
\end{equation}

On the other hand, in MCrcBK,
$\phi$ is obtained from the Fourier transform of
the numerical results of the running coupling BK (rcBK)
evolution equation~\cite{Albacete:2007yr,Albacete:2009fh,Albacete:2010sy}:
\begin{equation}
\phi_{A,B}(k,x,b)=\frac{C_F}{\alpha_s(k)\,(2\pi)^3}\int d^2{\bf r}\
e^{-i{\bf k}\cdot{\bf r}}\,\nabla^2_{\bf
r}\,\mathcal{N}_G(r,Y\!=\!\ln(x_0/x),b)\,.
\label{phi}
\end{equation}
where $C_F=(N_c^2-1)/2N_c$, $x_0=0.01$, and
$\mathcal{N}_G$ is related to the quark dipole scattering amplitude that solves
the rcBK equation, $\mathcal{N}$ as follows:
\begin{equation}
\mathcal{N}_G(r,x)=2\,\mathcal{N}(r,x)-\mathcal{N}^2(r,x)\,.
\end{equation}
In MC-KLN model, $x$ dependence is determined by eq.~(\ref{eq:qs2}),
but in rcBK, $x$ dependence can be obtained from the equation.
Therefore, we expect that MCrcBK model has more predictive power than
MC-KLN model.

\section{Results}

\subsection{MC-KLN}

\begin{figure}[htb]
\begin{minipage}{0.48\hsize}
%\begin{wrapfigure}{r}{0.5\textwidth}
\begin{center}
\includegraphics[width=7.8cm]{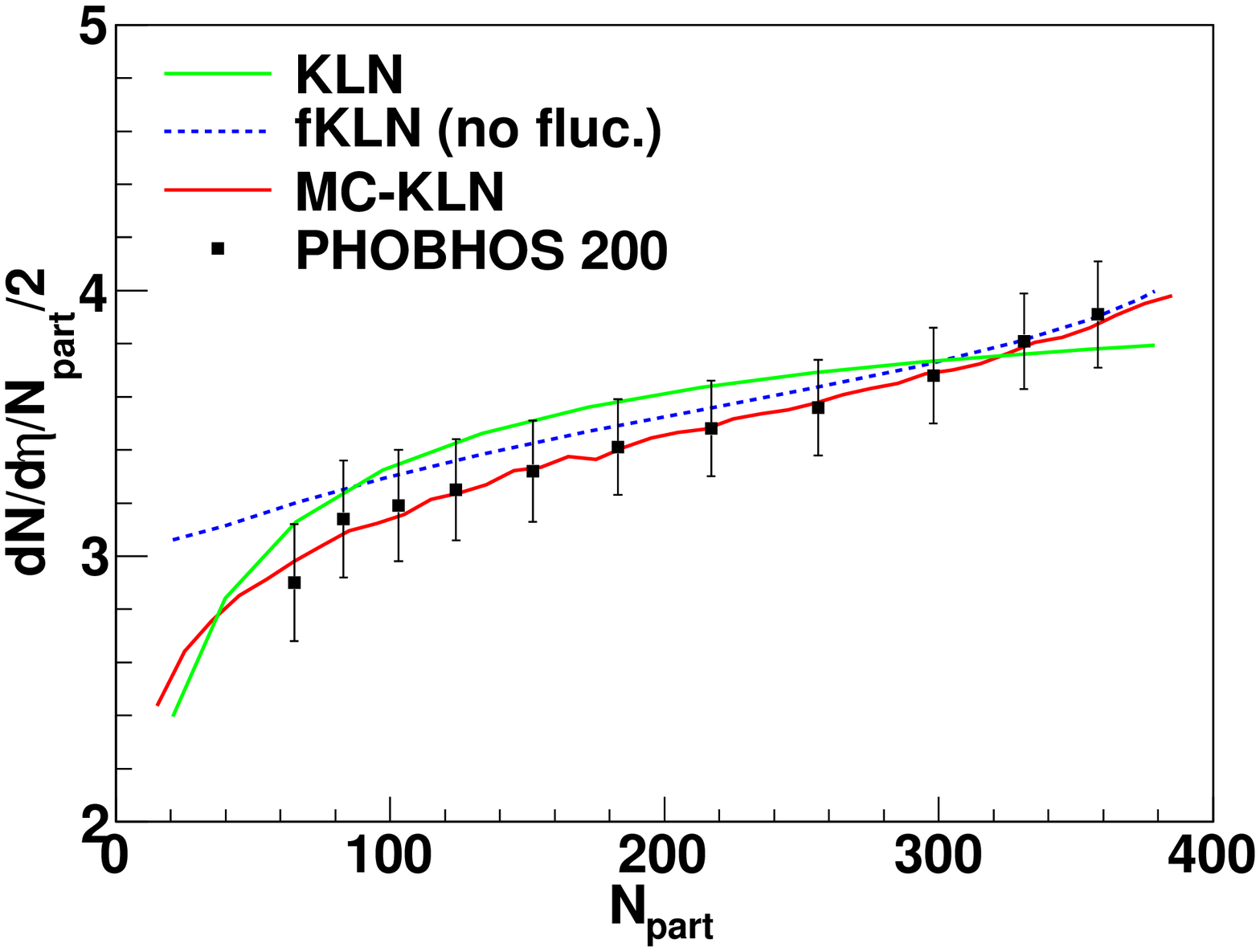}
\caption{\label{fig:nchnpart}
Centrality dependence of charged hadron multiplicity
in Au+Au collision
from the KLN, fKLN, and MC-KLN model
are compared with PHOBOS data \cite{phobos}
}
\end{center}
%\end{wrapfigure}
\end{minipage}
\hfill
%%%%%%%%%%%%%%%%%%%%%%%%%%%%%%%%%%%%%%%%%%%%%%%%%%%%%%%%%%%%%%%%%%%%%%%%%%%
\begin{minipage}{0.48\hsize}
%%%%%%%%%%%%%%%%%%%%%%%%%%% Fig. 2 %%%%%%%%%%%%%%%%%%%%%%%%%%%%%%%%%%%%%%%
%\begin{figure}[htb]
\begin{center}
\includegraphics[width=7.8cm]{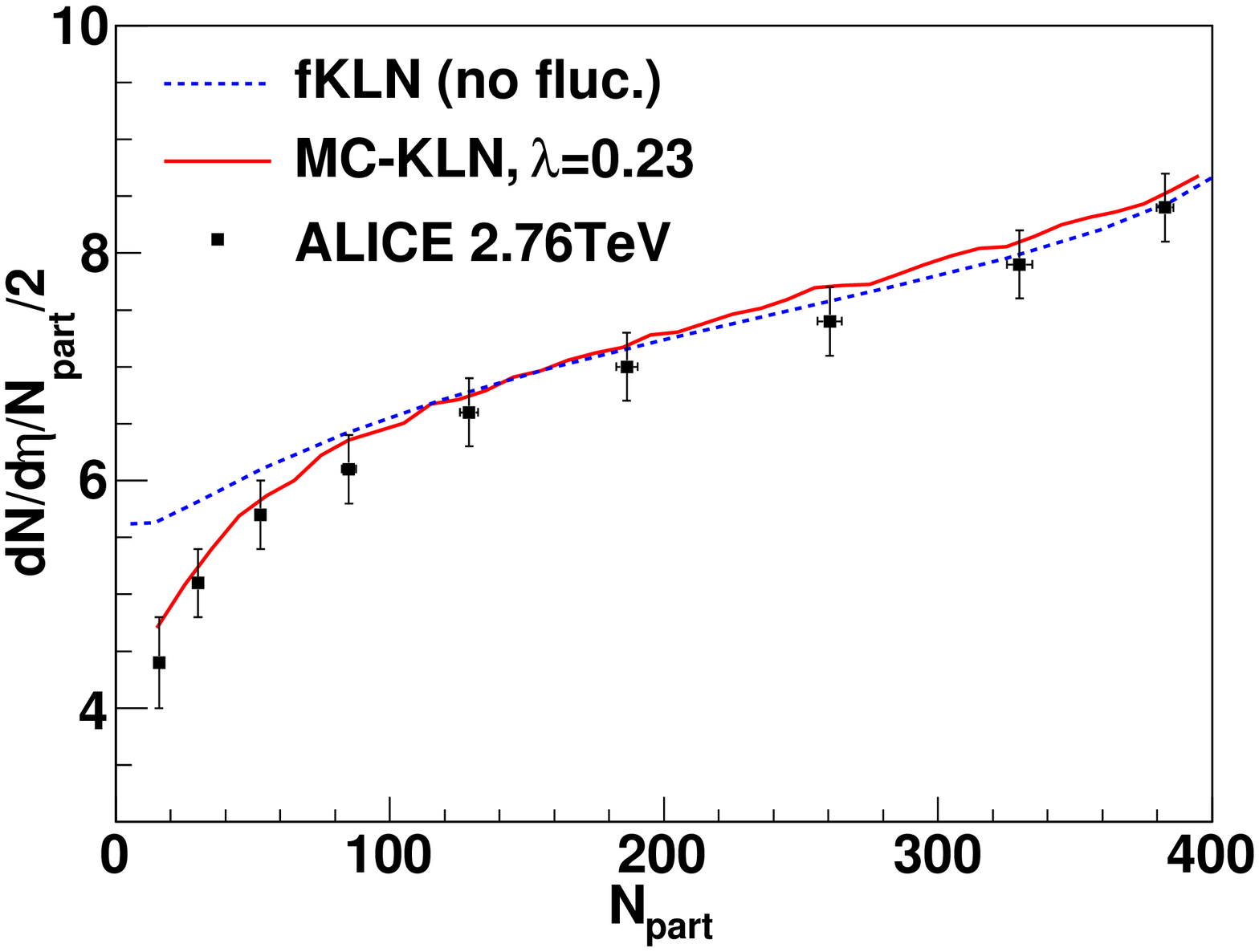}
\caption{
Centrality dependence of charged hadron multiplicity
in Pb+Pb collision at $\sqrt{s_{NN}} = 2.76$ TeV
from the fKLN, and MC-KLN model
are compared with ALICE data \cite{alice}
}
\label{fig:mcklnalice}
\end{center}
\end{minipage}
\end{figure}

In the figure~\ref{fig:nchnpart}, 
the charged particle multiplicities 
at RHIC and LHC
as a function of $N_\text{part}$
from KLN, fKLN, and MC-KLN model are compared to PHOBOS data.

%is compared with ALICE data \cite{Aamodt:2010pb,Aamodt:2010ft}.
%%%%%%%%%%%%%%%%%%%%%%%%%%% Fig. 1 %%%%%%%%%%%%%%%%%%%%%%%%%%%%%%%%%%%%%%%
%%%%%%%%%%%%%%%%%%%%%%%%%%%%%%%%%%%%%%%%%%%%%%%%%%%%%%%%%%%%%%%%%%%%%%%%%%%

In the KLN model, we plot the equation
\begin{equation}
\frac{1}{N_\text{part}}\frac{dN}{dy}
=c \ln \left(
   \frac{Q_s^2}{\Lambda^2_\text{QCD}}
       \right)\ .
\end{equation}
This equation is obtained by 
picking up the most important contribution of integration in 
$k_t$-factorized formula by using average nuclear saturation momentum.
On the other hand, fKLN model~\cite{fKLN}
does not take into account the effects of fluctuations
of nucleons inside a nucleus unlike the MC-KLN model.
The main difference between KLN and fKLN model is that
fKLN model uses the local saturation scale instead of assuming
a average nuclear saturation momentum as in KLN.
Therefore, we see that computations with local saturation momentum
improve results.
A comparison between fKLN and MC-KLN model shows the effects
of fluctuations of the position of nucleon inside a nucleus.
One sees such effect in the peripheral collision which
suggests that it is important to take into account fluctuation
effect in the discussion of the centrality dependence of 
particle multiplicities.

In Fig.~\ref{fig:mcklnalice},
the centrality dependence of charged hadron multiplicity in
Pb+Pb collision at $\sqrt{s_{NN}}=2.76$ TeV from ALICE experiment
is compared to the results from fKLN and MC-KLN.
Where we use $\lambda=0.23$ in Eq.~(\ref{eq:qs2})
which controls the $x$-evolution speed.
If $\lambda$ is small, multiplicity becomes small, and
we may think additional particle production mechanism during
the evolution of the system, especially before thermalization.

%%%%%%%%%%%%%%%%%%%%%%%%%%% Fig. 3 %%%%%%%%%%%%%%%%%%%%%%%%%%%%%%%%%%%%%%%
\begin{figure*}[htb]
\includegraphics[width=3.4in]{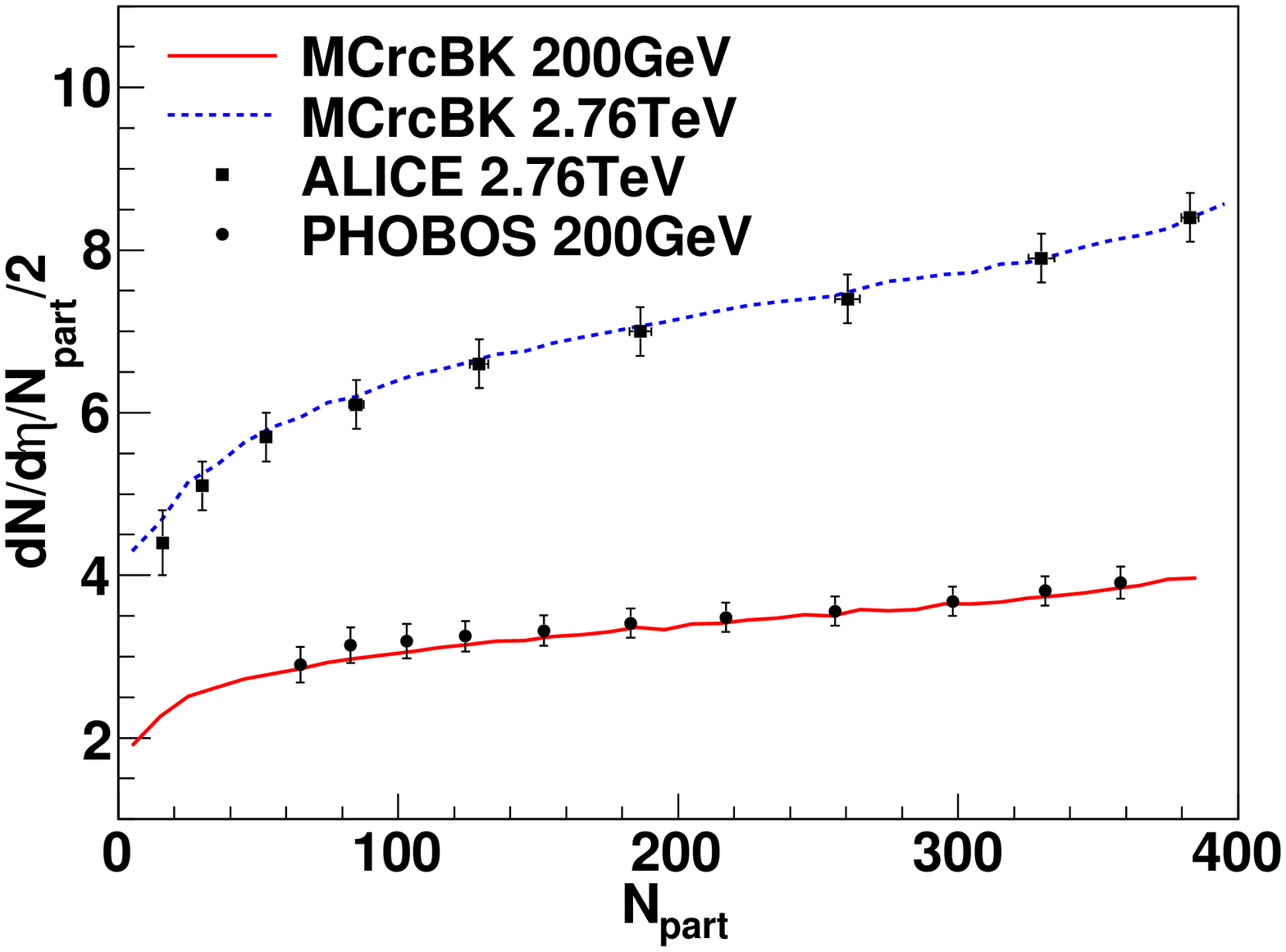}
\includegraphics[width=3.4in]{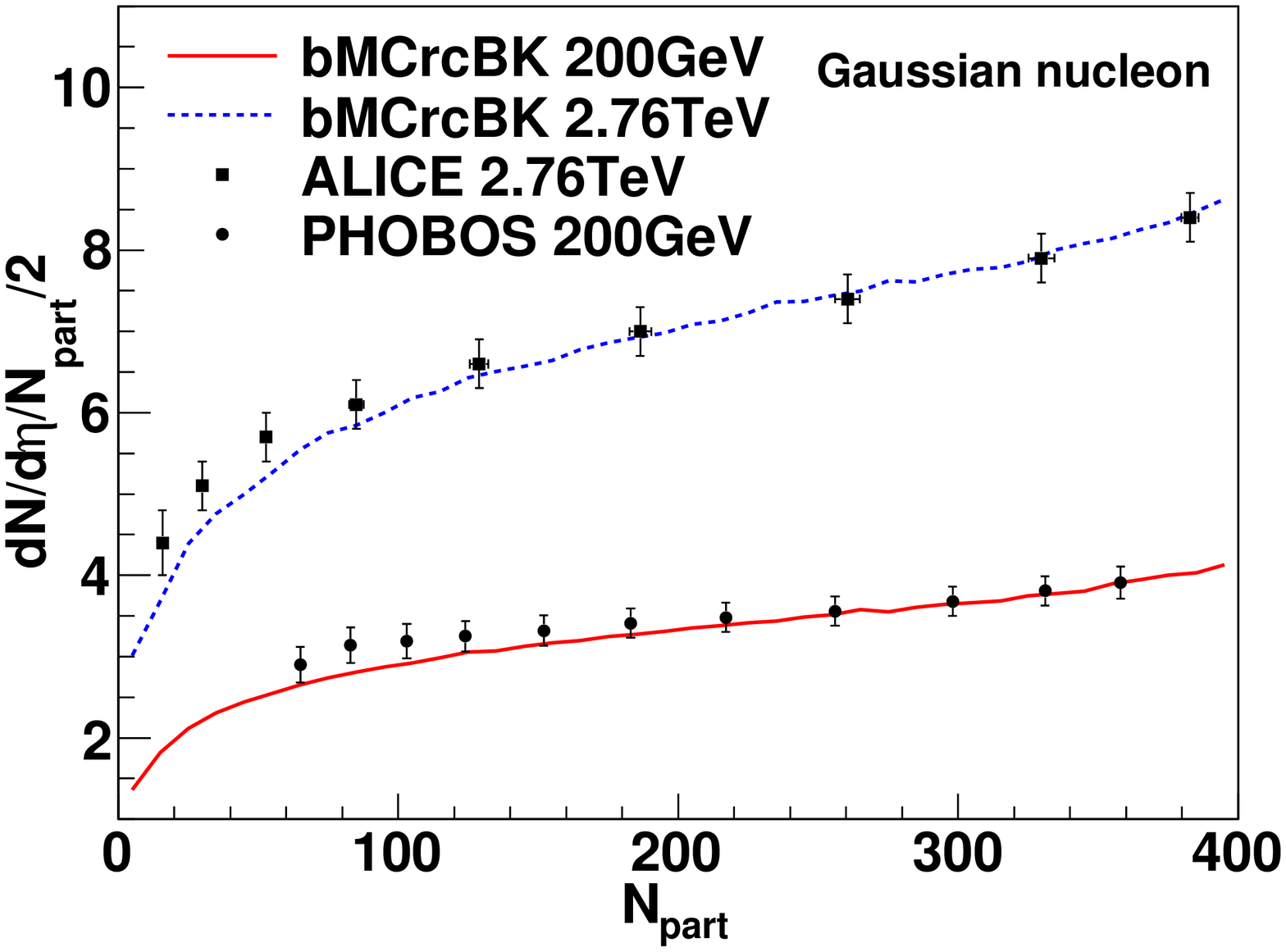}
\caption{Centrality dependence of charged hadron multiplicity
in Pb+Pb collision at $\sqrt{s_{NN}} = 2.76$ TeV
and Au+Au collision at $\sqrt{s_{NN}} = 200$ GeV
from the MCrcBK model are compared.
}
\label{fig:rcbk}
\end{figure*}
%%%%%%%%%%%%%%%%%%%%%%%%%%%%%%%%%%%%%%%%%%%%%%%%%%%%%%%%%%%%%%%%%%%%%%%%%%%

\subsection{MCrcBK}
We now compare in Fig.\ref{fig:rcbk}, 
the centrality dependence of charged hadron with the 
results obtained from the MCrcBK model in which unintegrated
gluon function is taken from the numerical solution of the rcBK equation.
The initial condition for the rcBK evolution is assumed to be
MV model:
\begin{equation}
\mathcal{N}(r,Y\!=\!0)=
1-\exp\left[-\frac{r^2\,Q_{s0}^2}{4}\,
  \ln\left(\frac{1}{\Lambda\,r}+e\right)\right]\ ,
\label{mv}
\end{equation}
with the initial saturation scale $Q_{s0}^2=0.2$ GeV$^2$
for nucleon
and $\Lambda=0.2$ GeV.
For the running coupling, we use
\begin{equation}
\alpha_s(r^2) = \frac{12\pi}{5\ln \left(
                           \frac{4C^2}{r^2\Lambda^2}
                      \right)}
\end{equation}
with $C=1$ in solving rcBK equation.
rcBK unintegrated gluon function with these parameter set
describes the centrality dependence of charged hadron multiplicity
for both RHIC and LHC which indicates that most of the gluon
is produced at the first impact of the nuclear collision.
However, we should check parameter dependence carefully.
For example, if one changes the parameter $C$ in the running coupling,
evolution speed changes.

In the bottom up scenario~\cite{bottomup},
the number of gluon produced just after the collision will
increase by a factor of $\alpha_s(Q_s^2)^{-2/5}$ during the subsequent
evolution of the system.
It would be interesting to include such effect into MCrcBK calculations.

\subsection{Gaussian shape}

So far we assumed that nucleus-nucleus collision is described by 
the incoherent sum of nucleon-nucleon collisions which will
occur when transverse distance squared between two nucleons is
smaller than the inelastic proton-proton cross section $\sigma_{NN}$
divided by $\pi$:
\begin{equation}
(x_i - x_j) ^2 + ( y_i - y_j)^2 \le 
\frac{\sigma_{NN}}{\pi}
\end{equation}
This amounts to assume that nucleon is hard sphere (or disk).
However, this approximation may not work in very high energy
hadronic collisions.
Let us think about the Gaussian shape nucleon in order to
take into account the effect of extension of nucleon size
as incident energy increases.
In this case the thickness function becomes
\begin{equation}
T_p(r) = \frac{1}{2\pi B}\exp[-r^2/(2B)]\ ,
\end{equation}
and the probability of nucleon-nucleon collision $P(b)$ at impact parameter $b$
is
\begin{equation}
P(b) = 1 - \exp[-k T_{pp}(b)],\qquad T_{pp}(b) = \int d^2s \, T_p(s)\,
T_p(s-b)\ .
\end{equation}
where (perturbatively) $k$ corresponds to the product of gluon-gluon
cross section
and gluon density squared.
We fix $k$ so that integral over impact parameter
becomes the nucleon-nucleon inelastic cross section $\sigma_{NN}$ at
the given energy:
\begin{equation}
\sigma_{NN} = \int d^2b \left(
   1-\exp[-kT_{pp}(b)]
    \right) .
\end{equation}
In this work, we use 
$B=0.2$ fm$^2$,
$\sigma_{NN}=41.94$ mb for $\sqrt{s_{NN}}=200$ GeV
and
$\sigma_{NN}=61.36$ mb for $\sqrt{s_{NN}}=2.76$ TeV.
The result of this model is plotted in right hand side of Fig.~\ref{fig:rcbk}.
Model underpredicts the multiplicity at peripheral collisions.
One possible interpretation may be the following:
viscosity is large at large impact parameter, entropy production
may be larger as impact parameter becomes large.

\subsection{Eccentricity and transverse energy}

\begin{figure}[htb]
\includegraphics[width=3.4in]{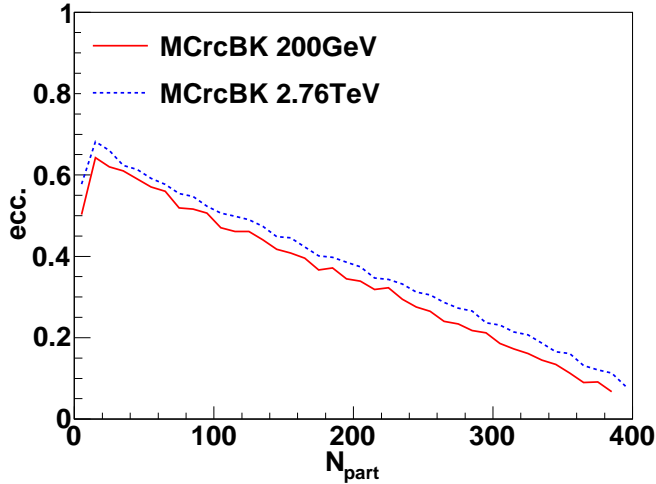}\hspace{2pc}%
\begin{minipage}[b]{14pc}
\caption{Centrality dependence of eccentricity
with respect to reaction plane in Pb+Pb collisions
at $\sqrt{s_{NN}}=$ 2.76 TeV and in Au+Au collisions
at $\sqrt{s_{NN}}=$ 200 GeV.}
 \label{fig:ecc}
 \end{minipage}
 \end{figure}

Finally, we plot eccentricity defined by
\begin{equation}
\epsilon = \frac{\langle y^2 - x^2 \rangle}{\langle y^2 + x^2 \rangle}
\end{equation}
in Fig.~\ref{fig:ecc}.
Since the eccentricity is proportional to the magnitude of elliptic flow,
it is important to know the initial value of this quantity.
One sees that the incident energy dependence of the eccentricity
is very small according to the results from MCrcBK.

%%%%%%%%%%%%%%%%%%%%%%%%%%% Fig. 4 %%%%%%%%%%%%%%%%%%%%%%%%%%%%%%%%%%%%%%%
%%%%%%%%%%%%%%%%%%%%%%%%%%%%%%%%%%%%%%%%%%%%%%%%%%%%%%%%%%%%%%%%%%%%%%%%%%%

%%%%%%%%%%%%%%%%%%%%%%%%%%% Fig. 5 %%%%%%%%%%%%%%%%%%%%%%%%%%%%%%%%%%%%%%%
\begin{figure*}[htb]
\includegraphics[width=3.2in]{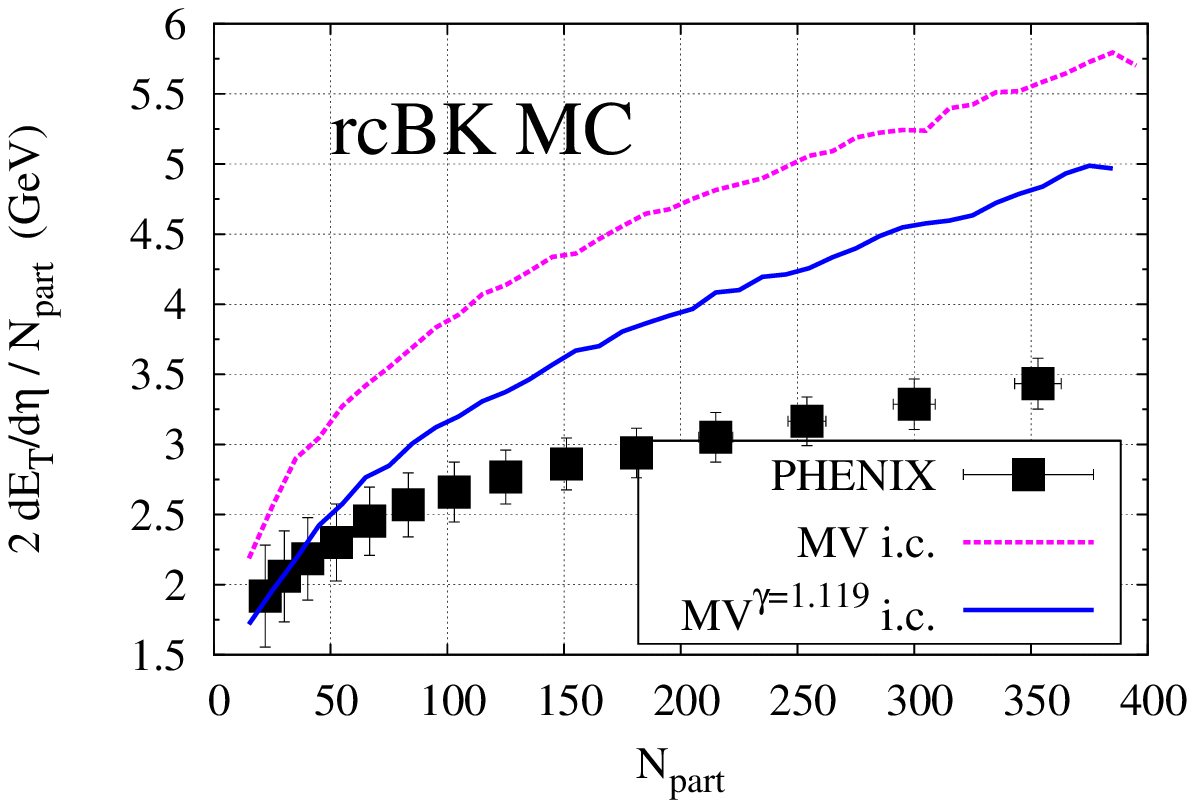}
\includegraphics[width=3.2in]{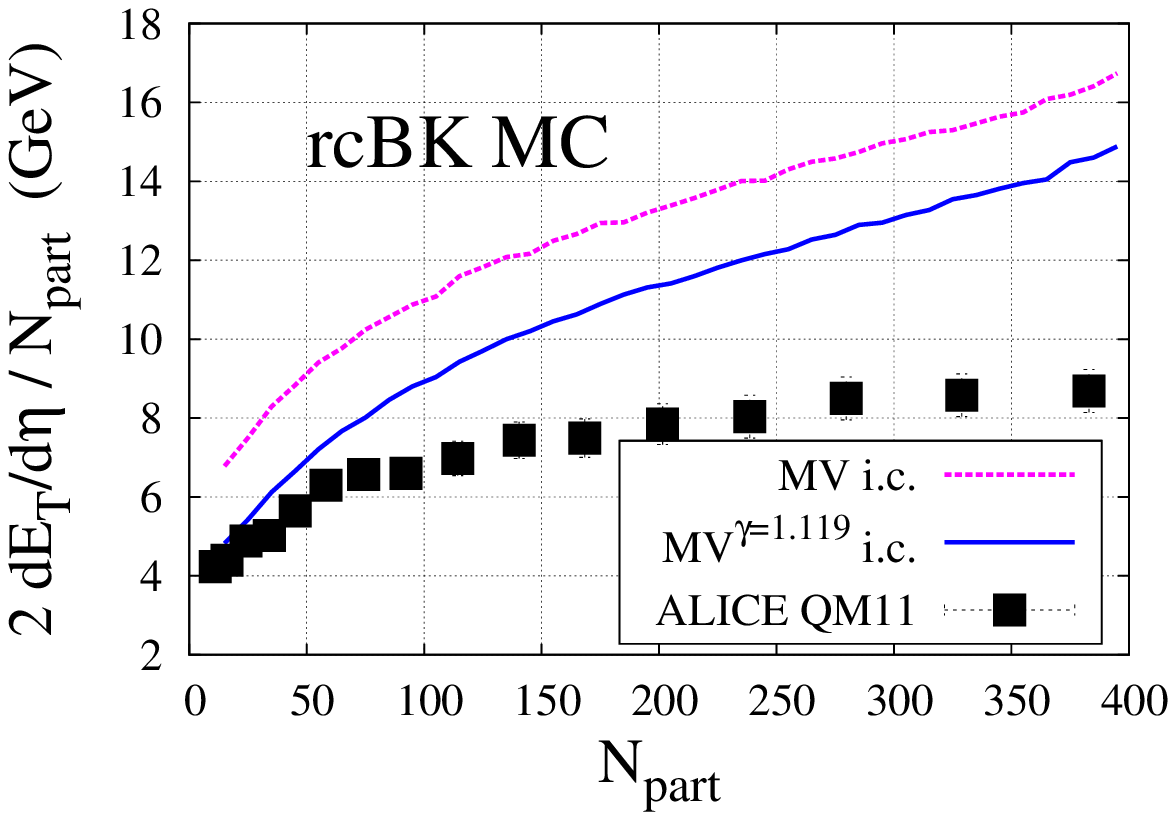}
\caption{Centrality dependence of the transverse energy
in Au+Au collision at $\sqrt{s_{NN}} = 200$ GeV (left) and
Pb+Pb collision at $\sqrt{s_{NN}} = 2.76$ TeV (right).
PHENIX data from \cite{PHENIX_Et}, ALICE data from \cite{ALICE_Et}.
}
\label{fig:rcbkEt}
\end{figure*}
%%%%%%%%%%%%%%%%%%%%%%%%%%%%%%%%%%%%%%%%%%%%%%%%%%%%%%%%%%%%%%%%%%%%%%%%%%%
Fig.~\ref{fig:rcbkEt} shows the centrality dependence of the transverse
energy at central rapidity for the MV initial condition, and for an
initial condition featuring a more rapid fall-off at $k_\perp>Q_s$;
see ref.~\cite{mckt} for details. We note that $\sim 2.5\%$ ($0.5\%$)
of the energy of the beams is predicted to be deposited initially into the
central rapidity region in central collisions. Longitudinal
hydrodynamic expansion ($-p\, \Delta V$ work) may reduce the
transverse energy by up to a factor of 2 \cite{Et_pdV}. In all, the
model based on rcBK evolution of the gluon distribution appears to be
consistent with the multiplicity and transverse energy data over a
good order of magnitude in both $N_{part}$ and $\sqrt{s_{NN}}$.

\section{Summary}
Within the MC-KLN and MCrcBK model, we compute the gluon
production based on the picture of Color Glass Condensate.
Both MC-KLN and MCrcBK model reproduce the centrality dependence
of charged hadron multiplicity at RHIC and LHC energies which
indicate that the entropy production during the thermalization
process may be small as well as the small viscosity after
thermalization.
However, we need detail systematic study of particle production
by looking at different observables together with the parameter
dependence.

\ack
The work of Y.N.\ was partly supported by
Grant-in-Aid for Scientific Research
No.~20540276.
A.D.\ acknowledges support by the DOE Office of Nuclear Physics
through Grant No.\ DE-FG02-09ER41620
and from The City University of New York through the PSC-CUNY Research
Award Program, grant 63382-0041.

%%%%%%%%%%%%%%%%%%%%%%%%  References %%%%%%%%%%%%%%%%%%%%%%%%%%%%%%%%%%%%%%%%%

\end{document}